\documentclass[12pt]{article}

\usepackage{amsfonts,amsmath,amssymb,accents,mathrsfs}
\usepackage[retainorgcmds]{IEEEtrantools}
\usepackage{multirow}
\usepackage{multicol}
\usepackage{tikz}
\usepackage{graphicx,color}
\usepackage{booktabs}
\usepackage{placeins}
\usepackage[colorlinks,linkcolor=Blue,citecolor=Blue,urlcolor=Blue,bookmarks,
bookmarksnumbered]{hyperref}
\usepackage[nosort]{cite}
\usepackage[scaled=0.85]{helvet}
\usepackage[T1]{fontenc}
\usepackage[utf8]{inputenc}
\usepackage{cancel}

\definecolor{Red}    {rgb}{0.90,0.00,0.12} 
\definecolor{Blue}   {rgb}{0.00,0.00,1.00} 
\definecolor{Green}  {rgb}{0.10,0.70,0.10} 
\definecolor{Turque} {rgb}{0.00,0.65,0.85} 
\definecolor{Orange} {rgb}{1.00,0.50,0.15} 
\definecolor{Magenta}{rgb}{1.00,0.00,1.00} 
\definecolor{Gold}   {rgb}{1.00,0.75,0.25} 
\definecolor{Seaweed}{rgb}{0.01,0.24,0.09} 
\definecolor{Purple} {rgb}{0.50,0.25,0.55} 
\definecolor{Brown}  {rgb}{0.43,0.26,0.32} 
\definecolor{grey1}  {rgb}{0.20,0.20,0.20} 
\definecolor{grey2}  {rgb}{0.40,0.40,0.40} 
\definecolor{grey3}  {rgb}{0.60,0.60,0.60} 
\definecolor{grey4}  {rgb}{0.80,0.80,0.80} 
\definecolor{grey5}  {rgb}{0.90,0.90,0.90} 

\def\a{{\alpha}}
\def\b{{\beta}}

\def\d{{\delta}}

\def\S{{\Sigma}}

\def\ad{{\dot{\alpha}}}
\def\bd{{\dot{\beta}}}

\def\N{{\mathcal{N}}}

\def\J{{\mathcal{J}}}
\def\T{{\mathcal{T}}}

\def\K{{\mathcal{K}}}
\def\W{{\mathcal{W}}}

\def\D{{\rm D}}
\def\Dd{{\bar{\rm D}}}
\def\pa{\partial}

\def\be{\begin{equation}}
\def\ee{\end{equation}}
\def\bea{\begin{IEEEeqnarray*}}
\def\eea{\end{IEEEeqnarray*}}
\def\n{\IEEEyesnumber}
\def\sn{\IEEEyessubnumber}

\makeatletter
\def\section{\@startsection{section}{1}{\z@}
              {3ex plus-1ex minus-.2ex}{1pt plus1pt}
              {\large\sf\bfseries\boldmath}}
\def\subsection{\@startsection{subsection}{2}{\z@}
              {1.5ex plus-1ex minus-.2ex}{0.01pt plus1pt}{\sf\slshape}}
\def\subsubsection{\@startsection{subsubsection}{3}{\z@}
              {1.5ex plus-1ex minus-.2ex}{0.01pt plus0.2pt}{\sf\boldmath}}
\def\paragraph{\@startsection{paragraph}{4}{\z@}
              {.75ex \@plus.5ex \@minus.2ex}{-2mm}{\sf\bfseries\boldmath}}
\makeatother

   \parskip\medskipamount     \lineskip=0pt
\thicklines
\begin{document}
\thispagestyle{empty}
\noindent{\small
\hfill{HET-1778 {~} \\ 
$~~~~~~~~~~~~~~~~~~~~~~~~~~~~~~~~~~~~~~~~~~~~~~~~~~~~~~~~~~~~$
$~~~~~~~~~~~~~~~~~~~~\,~~~~~~~~~~~~~~~~~~~~~~~~~\,~~~~~~~~~~~~~~~~$
 {~}
}
\vspace*{6mm}
\begin{center}
{\large \bf 
Integer superspin supercurrents of matter supermultiplets
\vspace{3ex}
} \\   [9mm] {\large { I. L.
Buchbinder,\footnote{joseph@tspu.edu.ru}$^{a,b}$ S.\ James Gates,
Jr.,\footnote{sylvester\_gates@brown.edu}$^{c}$ and K.\
Koutrolikos\footnote{konstantinos\_koutrolikos@brown.edu}$^{c}$ }}
\\*[8mm]
\emph{
\centering
$^a$Department of Theoretical Physics,Tomsk State Pedagogical University,\\
Tomsk 634041, Russia
\\[6pt]
$^b$National Research Tomsk State University,\\
Tomsk 634050, Russia
\\[6pt]
$^{c}$Department of Physics, Brown University,
\\[1pt]
Box 1843, 182 Hope Street, Barus \& Holley 545,
Providence, RI 02912, USA
}
 $$~~$$
  $$~~$$
 \\*[-8mm]
{ ABSTRACT}\\[4mm]
\parbox{142mm}{\parindent=2pc\indent\baselineskip=14pt plus1pt
In recent papers \cite{BGK1,KKvU} we demonstrated 
that consistent and non-trivial \emph{linear}
transformations of matter supermultiplets generate half-integer superspin
supercurrents and the cubic interactions between matter and half-integer
superspin supermultiplets. In this work we show that consistent and non-trivial
\emph{antilinear} transformations of matter superfields lead to the 
construction of integer superspin supercurrents and the cubic interactions 
between mater and integer superspin supermultiplets. Applying Noether's method 
to these transformations, we find new integer superspin supercurrents
for the case of a free massless chiral superfield. Furthermore, we use them to find new
integer superspin supercurrent multiplets for a massive chiral superfield 
and a chiral superfield with a linear superpotential. Also various selection 
rules for such interactions are found.  
}
\end{center}
$$~~$$
\vfill
\vfill
\clearpage
%
\section{Introduction}
\label{intro}
For non-supersymmetric theories there is a plethora of well-known results
on the topic of higher spin conserved currents
\cite{current1,current2,current3,current4,current5,current6,current7,current8,
current9,current10} and higher spin cubic interactions
\cite{lc4,lc5,lc6,lc7,lc9,lc10}. Recently some of these results
have been extended to supersymmetric theories. In 
a series of papers \cite{K1,BGK1,K2,K3,KKvU,BGK2,K4,BGK3} a variety of
supersymmetric, higher spin, currents have been constructed
for miscellaneous matter and higher spin supermultiplets while the 
corresponding
cubic interactions between matter and higher spin supermultiplets or between
higher spin and higher spin supermultiplets have been discussed.

In most of
these considerations, the multiplet of supercurrents were found by solving 
the
appropriate conservation equations. However, for \cite{BGK1,KKvU}
the foundation of the construction was the discovery of a \emph{linear}
non-trivial consistent, first order, higher spin transformation of 
matter superfields. Specifically, it was shown that the most general 
linear transformation of matter superfields, which is non-trivial and 
consistent with the various constraints of matter supermultiplets (chiral, or 
complex linear) is parametrized by terms that match the gauge symmetry of 
free, 
massless, half-integer superspin [$Y=s+1/2$] supermultiplets $(s+1,~s+1/2)$. 
The 
application of Noether's method to this kind of deformation lead us to the 
construction of higher spin supercurrents and higher spin supertraces which 
generate the cubic interactions of the various matter supermultiplets with 
the 
half-integer superspin supermultiplets. The construction is reminiscent of 
the 
way that linearized superdiffeorphisms lead to the construction of the 
supergravity supercurrent and supertrace of matter supermultiplets. 
Nevertheless, the absence of integer superspin [$Y=s$] supermultiplets 
($s+1/2,~s$) from the above consideration was intriguing. 

The purpose of this work is to find appropriate higher spin deformations of 
the 
matter superfields that lead via Noether's method to
the construction of integer superspin supercurrents and generate the 
cubic interactions with free, massless integer superspin supermultiplets.
We find that there exist non-trivial, \emph{antilinear} transformations
\footnote{A map $f:V\to W$ from one complex vector space $V$ to 
another $W$ is called antilinear if $f(au+bv)=a^*f(u)+b^*f(v)$ where $a,~b$ 
are 
complex numbers and $u,~v$ are elements of $V$. This is equivalent to a 
linear
map from $V$ to the complex conjugate vector space $\bar{W}$. 
The transformations we consider have this property, they are linear in the 
complex conjugate of the superfield.} of the matter superfields 
that will generate these interactions.
Specifically, we write the most general antilinear transformation
for a chiral superfield, demand it to be non-trivial and compatible
with the chiral constraint. The result is that parameters of the 
transformation
have the same structure with the gauge symmetry of free, massless integer 
superspin supermultiplets. That means by performing Noether's
method, we can construct the integer superspin supercurrent multiplet
(includes the supercurrent $\J_{\a(s)\ad(s-1)}$ and the supertrace
$\T_{\a(s-1)\ad(s-1)}$) and
generate the cubic interactions between the free, massless chiral superfield 
and 
the integer superspin supermultiplets ($s+1/2,s$). The results are
extended to the case of a free, massive chiral and a free chiral with linear
superpotential.

It is known that any $\N=1$ supersymmetric matter theory can be consistently 
coupled to supergravity with the help of the gravitational superfield. For 
that 
case the calculation of the
conserved supercurrent is straightforward.  One has to take the functional 
derivative of the interacting action with respect the gravitational 
superfield  
(see e.g. \cite{GGRS,BK}). However, this procedure is not applicable for 
higher spin 
theory  because we do not know the fully interacting theory at present.   The 
only alternative option we have is to follow
Noether's method in order to construct directly the higher spin supercurrent 
multiplet of the theory.

The paper is organized as follows. Section 2 reviews the philosophy and the
details of Noether's method as well as it provides to the non-expert reader 
the 
essentials for the description of $4D,~\N=1$ arbitrary integer superspin
supermultiplets for both the Poincar\'e and conformal cases. In section 3,
we consider first order transformations (in the spirit of Noether's method)
of the chiral superfield which are antilinear and demonstrate the fixing of 
their parameters by requiring them to be consistent with the chiral 
constraint 
of the superfield and non-trivial. Sections 4 and 5 consider the case of a
single, free, massless chiral superfield and derive the conformal
and Poincar\'e supercurrents respectively using the deformations of section 
3.
In section 6, we extend this results for the two case of a free, massive 
chiral 
superfield and a free chiral with linear superpotential. In the last section 
7,
we discuss and summarize our results. 

\section{Gauge invariant interacting theories of matter with gauge fields}
It is a fact of physics that a manifestly Lorentz invariant and local 
description of massless degrees of freedom with spin greater than $1/2$ 
requires 
the identification of various field configurations (gauge symmetries). As 
we transition perturbatively from free theories to interacting ones, the 
notion 
of this identification has to be re-examined in every step. This can be done 
systematically by expanding the action $S[\phi,h]$ and the transformation of 
all 
fields in a power series of a coupling constant $g$.
\bea{l}
S[\phi,h]=S_0[\phi]+gS_1[\phi,h]+g^2S_2[\phi,h]+\dots\n~,~\\
\delta \phi=0+g\delta_1\phi+g^2\delta_2\phi+\dots\n~,~\\
\delta h=\delta_0 h+g\delta_1 h+g^2\delta_2 h+\dots\n
\eea 
In the above expressions we consider the interaction of a set of matter 
fields
represented by $\phi$ with a set of gauge fields represented by $h$.
Matter fields do not have a zeroth order gauge transformation
($\d_{0}\phi=0$), whereas gauge fields do ($\d_{0} h\neq0$). The terms
$S_i[\phi,h]$ correspond to interaction terms of order $i+2$ in the
number of fields and $\delta_i$ is the part of transformation with terms
of order $i$ in the number of fields. The invariance of the theory under 
these
transformations can be studied iteratively, order by order. For cubic order 
terms $S_1[\phi,h]$ we get
\bea{l}\n\label{ic}
\int\left\{
\frac{\delta S_0}{\delta\phi}~\delta_1\phi
+\frac{\delta S_1}{\delta h}~\delta_0 h\right\}=0~.
\eea
The above expression is a symbolic one. There are a number of hidden
summations over `repeated fields'\footnote{There is a summation over
hidden external indices that count the number of matter and gauge fields,
there is a summation over the hidden spacetime indices of the gauge fields
and an integration over the spacetime coordinates.}
which are suppressed and represented by the integral sign. This invariance condition (up 
to
cubic terms) makes very clear the importance of the first order correction in 
the transformation of the matter fields, $\d_1\phi$. The starting action 
$S_0$ 
is known and the zeroth order transformations of gauge fields are also known. 
Hence, in order to find a \emph{consistent} set of \emph{non-trivial} cubic 
interactions $S_1[\phi,h]$ we must find a \emph{non-trivial} $\d_1\phi$. In 
this 
consideration, trivial interactions and trivial transformations are the ones 
that can be absorbed by an appropriate redefinition of the fields or in other 
words they vanish under the consideration of the equations of motion.

Cubic interactions of a matter theory with gauge fields can be written in the 
form $jh$ where $j$ is a current constructed out of the matter fields which 
plays the role of the source. For these types of interactions, condition
(\ref{ic}) takes the form
\bea{l}\n\label{icj}
\int\left\{
\frac{\delta S_0}{\delta\phi}~\delta_1\phi
+j~\delta_0 h\right\}=0
\eea
from which one can recover the conservation law of the current $j$ by using 
the
equations of motion (up to the appropriate order, for this case it is
$\frac{\delta S_0}{\delta\phi}=0$) and the structure of the gauge 
transformation 
of $h$ ($\d_{0}h=\pa\lambda$).

In recent papers \cite{BGK1,KKvU}, this approach has been used in order to 
construct conserved,
higher spin supercurrents for the chiral ($\Phi,~\Dd_{\ad}\Phi=0$) and
complex linear ($\S,~\Dd^2\S=0$) supermultiplets. In these papers we 
considered
the most general, non-trivial, first order transformations $\d_1\Phi,~\d_{1}
\S$
which depend linearly on $\Phi$ and $\S$ respectively. These transformations 
are
a higher spin extension of linearized superdiffeomorphism and like 
superdiffeomorphism generate interactions to supergravity supermultiplet 
$(2,3/2)$, they generate interactions to arbitrary higher spin 
supermultiplets
of type $(s+1,~s+1/2)$ (called half-integer superspin supermultiplets) for 
any
non-negative integer $s$.

In this work, we explore the possibility of non-trivial, first order
transformations that depend antilinearly on the
matter superfield. In the following sections we will find that such 
transformations do exist and generate interactions to arbitrary higher spin
supermultiplets of type $(s+1/2,~s)$ (called integer superspin 
supermultiplets).
We briefly remind the non-expert reader that the superspace Lagrangian
description of free, massless, super-Poincar{\'e}, arbitrary high ($s\ge2$), 
integer
superspin supermultiplet involves a fermionic superfield
$\Psi_{\a(s)\ad(s-1)}$ and a real bosonic superfield $V_{\a(s-1)\ad(s-1)}$ 
with
the following zeroth order gauge transformations
\bea{l}\n\label{isst}
\d_0\Psi_{\a(s)\ad(s-1)}=-\D^2L_{\a(s)\ad(s-1)}+\tfrac{1}{(s-1)!}~\Dd_{(\ad_{
s-1}}\Lambda_{\a(s)\ad(s-2))}~,\sn\label{PsiP}\vspace{1ex}\\  
\d_0 V_{\a(s-1)\ad(s-1)}=\D^{\a_s}L_{\a(s)\ad(s-1)}
+\Dd^{\ad_s}\bar{L}_{\a(s-1)\ad(s)}\sn~.
\eea
Off-shell, this supermultiplet carries $8s^2+8s+4$ bosonic and equal number 
of
fermionic degrees for freedom\footnote{A detailed counting of the off-shell
degrees of freedom can be found in \cite{hss6,hss8}.}. The
physical\footnote{The on-shell degrees of freedom are the 2 helicities of
spin $j=s+1/2$ and the two helicities of spin $j=s$} (propagating) degrees
of freedom are described by a field strength\footnote{This is the simplest
gauge invariant object that does not vanish on-shell.} superfield $
\W_{\a(2s)}$
\bea{l}
\W_{\a(2s)}\sim\Dd^2\D_{(\a_{2s}}\pa_{\a_{2s-1}}{}^{\ad_{s-1}}
\pa_{\a_{2s-2}}{}^{\ad_{s-2}}\dots\pa_{\a_{s+1}}{}^{\ad_{1}}\Psi_{\a(s)
\ad(s-1)}~.
\n\label{W}
\eea
The super-field strength is chiral $(\Dd_{\bd}\W_{\a(2s)}=0)$ and on-shell
satisfies the following equation of motion:
\bea{l}
\D^{\b}\W_{\b\a(2s-1)}=0~.\n
\eea
There is also a super-conformal integer superspin supermultiplet. Its 
Lagrangian description is given in terms of the super-field strength
$\W_{\a(2s)}$. Similarly with the super-Poincar\'e case, the super-field 
strength can be expressed in terms of a prepotential $\Psi_{\a(s)\ad(s-1)}$ 
(as in (\ref{W})) whose gauge transformation saturates the maximum symmetry 
group of 
$\W_{\a(2s)}$
\bea{l}\n\label{PsiC}
\d_0\Psi_{\a(s)\ad(s-1)}=\tfrac{1}{s!}~\D_{(\a_s}\Xi_{\a(s-1))\ad(s-1)}
+\tfrac{1}{(s-1)!}~\Dd_{(\ad_{s-1}}\Lambda_{\a(s)\ad(s-2))}~.
\eea
As is 
demonstrated in (\ref{icj}) the conservation law of the current (multiplet)
is determined by the gauge transformation of the gauge fields. Similarly, one 
can
use (\ref{isst}) and (\ref{PsiC}) to extract the corresponding superspace
conservation equations:
\bea{l}
\text{Poincar\'e:~~~}\D^2\J_{\a(s)\ad(s-1)}=\tfrac{1}{s!}~\D_{(\a_s}\T_{\a(s-1))\ad(s-1)}~~,~~
\Dd^{\ad_{s-1}}\J_{\a(s)\ad(s-1)}=0~~,\vspace{2ex}\\
\text{Conformal:~}\D^{\a_{s}}\J_{\a(s)\ad(s-1)}=0
~~,~~\Dd^{\a_{s-1}}\J_{\a(s)\ad(s-1)}=0~~.
\eea
These are the superspace conservation equations for the integer superspin
supercurrent $\J_{\a(s)\ad(s-1)}$ and supertrace $\T_{\a(s-1))\ad(s-1)}$
\footnote{The supertrace is relevant only to the super-Poincare 
higher spin supermultiplets because their description requires an additional
compensating superfield. For integer superspins this is the real
$V_{\a(s-1)\ad(s-1)}$ superfield.}
for the super-Poincar\'e and the super-conformal cases respectively.

For a detailed review of free, massless, supersymmetric higher spins we refer
the reader to the following.
The first Lagrangian description of supersymmetric, massless, higher spins in 
$4D$ 
Minkowski space 
was done in \cite{hss1,V}, using components with on-shell supersymmetry. A 
natural
approach to the off-shell formulation is to use the superspace and superfield 
methods (see e.g.\cite{GGRS,BK}). A superfield description of 
free supersymmetric massless, higher spin theories was presented for the first 
time 
in \cite{hss2,hss3,hss4} for both Minkowski and AdS spaces. This approach has
been further explored in \cite{BKS,GKS96a,GKS,hss5}.
Later studies of free supersymmetric, massless higher spin supermultiplets
include \cite{hss6,hss7,hss8}.

\section{Anti-linear transformation of the chiral superfield}
Let's consider a chiral superfield $\Phi$. The most general antilinear 
transformation one can write is
\footnote{We are following the ``\emph{Superspace}'' \cite{GGRS} conventions.}:
\bea{l}\n\label{dphi}
\d_1\Phi=\sum_{k=0}^{\infty}\left\{\vphantom{\frac12}\right.
~~A^{\a(k)\ad(k+1)}~\Dd_{\ad_{k+1}}\D_{\a_k}\Dd_{\ad_k}
\dots\D_{\a_1}\Dd_{\ad_1}\bar{\Phi}\vspace{-1ex}\\
\hspace{12ex}+~
\Delta^{\a(k)\ad(k)}~\D_{\a_k}\Dd_{\ad_k}\dots\D_{\a_1}\Dd_{\ad_1}\bar{\Phi}
\\
\hspace{12ex}+~
\Gamma^{\a(k+1)\ad(k)}~\D_{\a_{k+1}}\Dd^2\D_{\a_k}\Dd_{\ad_k}
\dots\D_{\a_1}\Dd_{\ad_1}\bar{\Phi}\\
\hspace{12ex}+~
E^{\a(k)\ad(k)}~\Dd^2\D_{\a_k}\Dd_{\ad_k}\dots\D_{\a_1}\Dd_{\ad_1}\bar{\Phi}
\left.\vphantom{\frac12}\right\}~.
\eea
The consistency of this transformation with the chiral condition of $\Phi$, 
$\Dd_{\ad}\Phi=0$ constraints the parameters of the transformation in the
following way:
\bea{l}\n
\Dd_{\bd}A^{\a(k)\ad(k+1)}
+\tfrac{1}{(k+1)!}~\Delta^{\a(k)(\ad(k)}~\delta_{\bd}{}^{\ad_{k+1})}=0
\sn~,\vspace{1ex}\\
\Dd_{\bd}\Delta^{\a(k)\ad(k)}=0\sn~,\vspace{1ex}\\
\tfrac{k+1}{k+2}~\Delta^{\a(k+1)\ad(k+1)}~C_{\bd\ad_{k+1}}
+\Dd_{\bd}\Gamma^{\a(k+1)\ad(k)}=0\sn~,\vspace{1ex}\\
A^{\a(k+1)\ad(k+2)}~C_{\bd\ad_{k+2}}+\Dd_{\bd}E^{\a(k+1)\ad(k+1)}
-\tfrac{1}{(k+1)!}~\Gamma^{\a(k+1)(\ad(k)}~\delta_{\bd}{}^{\ad_{k+1})}=0
\sn~,\vspace{1ex}\\
A^{\ad}~C_{\bd\ad}+\Dd_{\bd}E=0\sn~.
\eea
The solution of the above set of constraints is:
\bea{l}\n\label{sol}
A_{\a(k)\ad(k+1)}=\tfrac{1}{(k+1)!}~\Dd_{(\ad_{k+1}}\bar{\xi}_{\a(k)
\ad(k))}
-~\Dd^2\bar{\ell}_{\a(k)\ad(k+1)}\sn\label{A}~,\vspace{1ex}\\
\Delta_{\a(k)\ad(k)}=\Dd^2\bar{\xi}_{\a(k)\ad(k)}\sn~,\vspace{1ex}\\
\Gamma_{\a(k+1)\ad(k)}=\tfrac{k+1}{k+2}~
\Dd^{\ad_{k+1}}\bar{\xi}_{\a(k+1\ad(k+1)}\sn\label{G}~,\vspace{1ex}\\
E_{\a(k)\ad(k)}=-\bar{\xi}_{\a(k)\ad(k)}
-\Dd^{\ad_{k+1}}\bar{\ell}_{\a(k)\ad(k+1)}\sn\label{E}
\eea
where $\bar{\xi}_{\a(k)\ad(k)}$ and $\bar{\ell}_{\a(k)\ad(k+1)}$ are 
arbitrary,
unconstrained superfields. An observation is that all the 
$\bar{\ell}_{\a(k)\ad(k+1)}$ terms in (\ref{sol}) can be adsorbed by 
doing a field redefinition of
$\bar{\xi}_{\a(k)\ad(k)}$ (
$\bar{\xi}_{\a(k)\ad(k)}~\to~\bar{\xi}_{\a(k)\ad(k)}
-\Dd^{\ad_{k+1}}\bar{\ell}_{\a(k)\ad(k+1)}$). However, because the 
$\bar{\xi}_{\a(k)\ad(k)}$ and the $\bar{\ell}_{\a(k)\ad(k+1)}$ parts
of (\ref{A}) are similar in structure to the terms that appear in the 
complex 
conjugate versions of the conformal (\ref{PsiC}) and Poincar\'e 
(\ref{PsiP})
transformations respectively, it will be convenient to consider them 
separately.
The significance of these two cases is related to the fact that
conformal integer superspin multiplets have a bigger gauge symmetry
(\ref{PsiC}) than the Poincar\'e integer superspin multiplets
(\ref{PsiP}). Only by further
constraining the gauge parameter $\Xi_{\a(s-1)\ad(s-1)}$ of the conformal
case down to $\Xi_{\a(s-1)\ad(s-1)}=\D^{\a_s}L_{\a(s-1)\ad(s-1)}$
one can introduce an appropriate compensating superfield that will
break the conformal symmetry to its Poincar\'e subgroup. It is interesting 
that this mismatch between the corresponding 
gauge transformations of conformal and Poincar\'e higher superspin 
multiplets appears only for the integer superspin values.
In \cite{GS32} it was demonstrated that for the matter 
gravitino supermultiplet [$Y=1$] ($3/2,1$) one can relax the Poincar\'e
gauge transformation to match the conformal one, by adding another 
compensating 
superfield with an algebraic (no derivatives) transformation law. Recently
\cite{K3} this mechanism was applied to higher integer superspin 
supermultiplets. However, this description is non-economical (requires
more superfields than it is necessary) and one can always use the 
algebraic nature of the transformation of the additional compensator in 
order to remove it. Hence, we will work using the (\ref{isst}) description 
where the transformations of the conformal and Poincar\'e supermultiplets 
differ. For the half-integer superspin multiplets 
the corresponding gauge transformations between conformal and Poincar\'e
cases are identical. This is can also be seen as the reason why the 
equivalent analysis for the cubic interaction of the chiral with the
half-integer superspin multiplets 
\cite{BGK1,KKvU} gave a unigue class 
of consistent transformations for the chiral superfield.

The most encouraging observation is the similarity between the
structure of the gauge parameters that parametrize the transformation of 
the chiral superfield and the gauge transformations of the conformal or 
Poincar\'e integer superspin supermultiplets. This is a hint that indeed 
we can find cubic interactions of the chiral superfield with the integer 
superspins.

Therefore, based on the above results we will consider the following two classes of chiral transformations:
\bea{l}\n\label{Edphi}
1.~\d_1\Phi=\sum_{k=0}^{\infty}\left\{\vphantom{\frac12}\right.
\sn\label{EdphiC}
\tfrac{1}{(k+1)!}~\Dd^{(\ad_{k+1}}\bar{\xi}^{\a(k)\ad(k))}~
\Dd_{\ad_{k+1}}
\D_{\a_k}\Dd_{\ad_k}\dots\D_{\a_1}\Dd_{\ad_1}\bar{\Phi}\\
\hspace{13ex}
~+~
\Dd^2\bar{\xi}^{\a(k)\ad(k)}~
\D_{\a_k}\Dd_{\ad_k}\dots\D_{\a_1}\Dd_{\ad_1}\bar{\Phi}\\
\hspace{13ex}~-~
\tfrac{k+1}{k+2}~
\Dd_{\ad_{k+1}}\bar{\xi}^{\a(k+1\ad(k+1)}~
\D_{\a_{k+1}}\Dd^2\D_{\a_k}\Dd_{\ad_k}
\dots\D_{\a_1}\Dd_{\ad_1}\bar{\Phi}\\
\hspace{13ex}
~-~
\bar{\xi}^{\a(k)\ad(k)}~\Dd^2\D_{\a_k}\Dd_{\ad_k}\dots\D_{\a_1}\Dd_{\ad_1}
\bar{\Phi}
\left.\vphantom{\frac12}\right\}\vspace{1ex}\\
2.~
\d_1\Phi=\sum_{k=0}^{\infty}\left\{\vphantom{\frac12}\right.
\sn\label{EdphiP}
\Dd^2\bar{\ell}^{\a(k)\ad(k+1)}~\Dd_{\ad_{k+1}}\D_{\a_k}\Dd_{\ad_k}\dots
\D_{\a_1}\Dd_{\ad_1}\bar{\Phi}\\
\hspace{13ex}~+~
\Dd_{\ad_{k+1}}\bar{\ell}^{\a(k)\ad(k+1)}~
\Dd^2\D_{\a_k}\Dd_{\ad_k}\dots\D_{\a_1}\Dd_{\ad_1}\bar{\Phi}
\left.\vphantom{\frac12}\right\}
\eea
We will demonstrate that the first will generate the conformal integer
superspin supercurrents and the corresponding interactions with the 
conformal integer superspin supermultiplets, whereas the second will lead 
to the Poincar\'e integer superspin supercurrents and the interactions of 
matter with Poincar\'e integer superspin supermultiplets.

\section{Conformal integer superspin supercurrents for free massless 
chiral}
Let's consider the case of a single, free, massless chiral superfield
\bea{l}
S_0=\int d^8z ~~\bar{\Phi}\Phi~~.\n\label{S0}
\eea
Using (\ref{Edphi}) and the action above, we can follow the steps of
Noether's method, in order to find the corresponding conserved 
supercurrents. These type of calculations are done
on-shell (modulo terms which depend on the equations of motion) where 
the conservation of the (super)current is revealed. In this case the
equation of motion takes the form $\Dd^2\bar{\Phi}=0$, hence in the
variation of the action we can ignore
terms that depend on $\Dd^2\bar{\Phi}$. Notice that our transformations 
(\ref{Edphi}) have a few terms of these type, hence their contribution
to the variation of the action can be immediately ignored. The same 
conclusion can be reached by a different argument which is based on 
the distinction of the various terms that appear in (\ref{Edphi}) into
two classes. The first one is the class of terms that will generate
non-trivial effects. The second one is the class of terms that have 
trivial contributions which can be absorbed by appropriate field redefinitions.
The parts of transformations (\ref{Edphi}) that depend on 
$\Dd^2\bar{\Phi}$ fall into the second class. As an example, consider the
$k=0$ part of (\ref{Edphi}) $(-\bar{\xi}+\Dd_{\ad}\bar{\ell}^{\ad})\Dd^2\bar{\Phi}$
and calculate its effect in the transformed action\footnote{Keeping only terms linear in $g$}:
\bea{l}
S_{0}=\int \bar{\Phi}\Phi~+~g\int\left[\vphantom{\frac12}
\bar{\Phi}~(-\bar{\xi}+\Dd_{\ad}\bar{\ell}^{\ad})\Dd^2\bar{\Phi}+c.c.
\right]~.
\eea
By doing an integration by part this can be written in the following form
\bea{l}
S_{0}=\int \bar{\Phi}\Phi~+~g\int\left[\vphantom{\frac12}
\Dd^2\left\{\bar{\Phi}~(-\bar{\xi}+\Dd_{\ad}\bar{\ell}^{\ad})\right\}~\bar{\Phi}+c.c.
\right]
\eea
hence, one can do the following redefinition 
$\Phi\to\Phi-g\Dd^2\left[\bar{\Phi}(-\bar{\xi}+\Dd_{\ad}\bar{\ell}^{\ad})
\right]$ and completely absorb the second term, up to order $g$. Similar 
arguments holds for all terms of (\ref{Edphi}) that depend on $
\Dd^2\bar{\Phi}$. Their effect in the variation of the action can be counteracted (up to order $g$) by redefinitions of the type
$\Phi\to\Phi-g\Dd^2\left\{\bar{\Phi}~\sum_{k=0}^{\infty}[\pa^{(k)}F_{\a(k)\ad(k)}
+\D^{\a_{k+1}}\pa^{(k)}G_{\a(k+1)\a(k)}]\right\}$ for arbitrary superfields $F_{\a(k)\ad(k)}$ and $G_{\a(k+1)\ad(k)}$.
These terms are precisely the 
terms that are dropped by following the ``on-shell'' approach, so the two
arguments are in complete agreement. With that in mind, we use (\ref{EdphiC}) to calculate the variation of
$S_{0}$ to be:
\bea{l}\n\label{dS}
\d_{g}S_{0}=g\sum_{k=0}^{\infty}\int\left[\vphantom{\frac12}
\tfrac{1}{(k+1)!}\D^{(\a_{k+1}}\xi^{\a(k))\ad(k)}
~\left\{(-i)^k\Phi~\pa^{(k)}\D\Phi\right\}
~+~
\D^2\xi^{\a(k))\ad(k)}
~\left\{(-i)^k\Phi~\pa^{(k)}\Phi\right\}
\right]+c.c.~~
\eea
The quantities inside the curly brackets are not uniquely defined because one
can consider improvement terms $A_{\a(k+1)\ad(k)}$ and $B_{\a(k)\ad(k)}$
that satisfy:
\bea{l}\n
\D^{\a_{k+1}}A_{\a(k+1)\ad(k)}=\D^2 B_{\a(k)\ad(k)}~~(\text{up to terms that depend on e.o.m}).
\eea
A general expression for the improvement terms is
\bea{l}\n
A_{\a(k+1)\ad(k)}=
\tfrac{k+1}{(k+2)!}~\D_{(\a_{k+1}}\zeta_{\a(k))\ad(k)}~+~
\tfrac{1}{k!}\Dd_{(\ad_{k}}\D^2\kappa_{\a(k+1)\ad(k-1))}~+~
X_{\a(k+1)\ad(k)}~~,\sn\vspace{0.8ex}\\
B_{\a(k)\ad(k)}= \zeta_{\a(k)\ad(k)}~+~
\tfrac{1}{k!}\Dd_{(\ad_k}\D^{\a_{k+1}}\kappa_{\a(k+1)\ad(k-1))}~+~
Y_{\a(k)\ad(k)}\sn
\eea
where $\D^{\a_{k+1}}X_{\a(k+1)\ad(k)}$=0 and $\D^2Y_{\a(k)\ad(k)}$=0 modulo terms that depend on $\Dd^2\bar{\Phi}$.
The superfield $X_{\a(k+1)\ad(k)}$ may include terms like
$\D^{\a_{k+2}}P^{(1)}_{\a(k+2)\ad(k)}$ or $\D^2P^{(2)}_{\a(k+1)\ad(k)}$
which identically satisfy $X$'s constraint due to the algebra of the
covariant spinorial derivatives. However, it is important to state that
there can be non-trivial solutions which do not fit into this form.
An example of this has been demonstrated in \cite{BGK2}.
A similar statement holds true for superfield $Y_{\a(k)\ad(k)}$.
Therefore, equation (\ref{dS}) can be written in the following way
\bea{l}\n
\d_{g}S_{0}=g\sum_{k=0}^{\infty}\int\left[\vphantom{\frac12}
\tfrac{1}{(k+1)!}\D^{(\a_{k+1}}\xi^{\a(k))\ad(k)}
~\J_{\a(k+1)\ad(k)}
~+~
\D^2\xi^{\a(k))\ad(k)}
~\T_{\a(k)\ad(k)}
\right]+c.c.~~
\eea
where
\bea{l}\n
\J_{\a(k+1)\ad(k)}=(-i)^k\Phi~\pa^{(k)}\D\Phi~+~
\tfrac{k+1}{(k+2)!}~\D_{(\a_{k+1}}\zeta_{\a(k))\ad(k)}~+~
\tfrac{1}{k!}\Dd_{(\ad_{k}}\D^2\kappa_{\a(k+1)\ad(k-1))}~+~
X_{\a(k+1)\ad(k)}~,~~~~~~~\sn\vspace{1.1ex}\\
\T_{\a(k)\ad(k)}=(-i)^k\Phi~\pa^{(k)}\Phi~+~
\zeta_{\a(k)\ad(k)}~+~
\tfrac{1}{k!}\Dd_{(\ad_k}\D^{\a_{k+1}}\kappa_{\a(k+1)\ad(k-1))}~+~
Y_{\a(k)\ad(k)}~.\sn
\eea
Exploiting the freedom of the unconstrained $\zeta_{\a(k)\ad(k)}$ improvement
term we can select it appropriately such that $\T_{\a(k)\ad(k)}$=0.
With this choice, the variation of $S_0$ reduces to:
\bea{l}\n\label{dSpreC}
\d_{g}S_{0}=g\sum_{k=0}^{\infty}\int\left[\vphantom{\frac12}
\tfrac{1}{(k+1)!}~\D^{(\a_{k+1}}\xi^{\a(k))\ad(k)}
~\J_{\a(k+1)\ad(k)}
\right]+c.c.~~
\eea
with 
\bea{l}
\J_{\a(k+1)\ad(k)}=\tfrac{(-i)^k}{k+2}~\Phi~\pa^{(k)}\D\Phi~-~
\tfrac{k+1}{k+2}(-i)^k~\D\Phi~\pa^{(k)}\Phi~+~
X_{\a(k+1)\ad(k)}
~-~\tfrac{k+1}{(k+2)!}~\D_{(\a_{k+1}}Y_{\a(k))\ad(k)}\n\label{J}
\vspace{1.0ex}\\
\hspace{12ex}
~+
\tfrac{1}{k!}~\Dd_{(\ad_{k}}\D^2\kappa_{\a(k+1)\ad(k-1))}
~-~
\tfrac{k+1}{k+2}\tfrac{1}{(k+1)!k!}~
\D_{(\a_{k+1}}\Dd_{(\ad_{k}}\D^{\b}\kappa_{\b\a(k))\ad(k-1))}~.
\eea
The $Y_{\a(k)\ad(k)}$ and $\kappa_{\a(k+1)\ad(k)}$ terms of (\ref{J}) can be
absorbed by appropriate redefinition of $X_{\a(k+1)\ad(k)}$ (they are consistent with
$\D^{\a_{k+1}}X_{\a(k+1)\ad(k)}$=0), hence we can simplify the expression for
$\J_{\a(k+1)\ad(k)}$:
\bea{l}
\J_{\a(k+1)\ad(k)}=\tfrac{(-i)^k}{k+2}~\Phi~\pa^{(k)}\D\Phi~-~
\tfrac{k+1}{k+2}(-i)^k~\D\Phi~\pa^{(k)}\Phi~+~
X_{\a(k+1)\ad(k)}~.\n\label{Jcurrent}\n
\eea
In order to get consistent 
interactions with conformal integer superspin supermultiplets 
($\Psi_{\a(s)\ad(s-1)}$) we have to consider the full transformation 
(\ref{PsiC}) and not just a part of it as it appears in (\ref{dSpreC}).
However we can write\footnote{Notice that we ignored the $k=0$ term, because it
does not correspond to higher spin supermultiplets ($k\geq1$) but to the matter 
gravitino supermultiplet. Although the analysis will go through even in that case,
for simplicity we will not include it.}
\bea{l}\n\label{dSC}
\d_{g}S_{0}=g\sum_{k=1}^{\infty}\int\left[\vphantom{\frac12}
\left\{~\tfrac{1}{(k+1)!}~\D^{(\a_{k+1}}\xi^{\a(k))\ad(k)}
~+~\tfrac{1}{k!}~\Dd^{(\ad_k}\lambda^{\a(k+1)\ad(k-1))}~\right\}
~\J_{\a(k+1)\ad(k)}
\right]+c.c.~~
\eea
if and only if $\J_{\a(k+1)\ad(k)}$ has the property 
$\Dd^{\ad_k}\J_{\a(k+1)\ad(k)}$=0, identically. An encouraging observation 
towards this direction is that the first term in (\ref{Jcurrent}) has this 
property. However, the last two terms of (\ref{Jcurrent}) do not comply for
generic $X_{\a(k+1)\ad(k)}$. This is reasonable because both these terms 
originated from the improvement terms consideration and include a lot of 
freedom. Hence, we must choose the improvement term $X_{\a(k+1)\ad(k)}$
appropriately such that
\bea{l}\n\label{Xconst}
\D^{\a_{k+1}}X_{\a(k+1)\ad(k)}=0~~(\text{up to}~\Dd^2\bar{\Phi}~\text{terms})~,
\sn\vspace{1ex}\\
\Dd^{\ad_k}\left[X_{\a(k+1)\ad(k)}
~-~\tfrac{k+1}{k+2}(-i)^k~\D\Phi~\pa^{(k)}\Phi\right]=0~~
(\text{identically})~.\sn
\eea
These two conditions will uniquely fix the improvement term $X_{\a(k+1)\ad(k)}$. 
To find the explicit expression of $X_{\a(k+1)\ad(k)}$, let's consider the
ansatz
\bea{l}
X_{\a(k+1)\ad(k)}=\sum_{p=0}^{k}~c_{p}~\pa^{(p)}\D\Phi~\pa^{(k-p)}\Phi~~.\n
\eea
Constraints (\ref{Xconst}) are equivalent to:
\vspace{-1ex}
\bea{l}\n
c_{k-p}=-\tfrac{p+1}{k-p+1}~c_{p}~~,~~p=0,1,\dots,k\sn\vspace{0.8ex}\\
c_{k-p-1}=\tfrac{k-p}{p+1}~c_{p}~~,~~p=1,2,\dots,k-2\sn\vspace{0.8ex}\\
c_{k-1}=k~c_0~-~\tfrac{k(k+1)}{k+2}~(-i)^k\sn
\eea
It is straightforward to prove that this system of recursive equations has a solution only for odd values of $k$ (~$k=2l+1~,~l=0,1,2,\dots$~)
\bea{l}\n
c_{k}=c_{0}=0~,\vspace{1ex}\sn\\
c_{p}=-\frac{(-i)^k}{k+2}~(-1)^p~\binom{k}{p}~\binom{k+1}{p+1}
~~,~~p=1,2,\dots ,k-1~~,~~k=2l+1~~.\sn
\eea
To complete the procedure and get an invariant action we must add to the
starting action the following cubic interactions term
\bea{l}
S_{I}=-g\sum_{l=0}^{\infty}\int~\Psi^{\a(2l+2)\ad(2l+1)}~\J_{\a(2l+2)\ad(2l+1)}
~+~c.c.\n\label{cici}
\eea
where the superfield $\Psi_{\a(2l+2)\ad(2l+1)}$ has a transformation that is 
dictated by the curly brackets of (\ref{dSC}). This transformation is of the 
same kind as (\ref{PsiC}) and therefore the superfield
$\Psi_{\a(2l+2)\ad(2l+1)}$ unambiguously describes the conformal integer 
superspin $Y=2l+2~(l=0,1,\dots)$ supermultiplet. The cubic interaction term 
is generated by the integer superspin supercurrent 
$\J_{\a(2l+2)\ad(2l+1)}$:
\bea{l}\n\label{Jspc}
\J_{\a(2l+2)\ad(2l+1)}=\frac{i(-1)^l}{2l+3}~\sum_{p=0}^{2l+1}~
(-1)^{p}~\binom{2l+1}{p}~\binom{2l+2}{p+1}~
\pa^{(p)}\D\Phi~\pa^{(2l+1-p)}\Phi~.
\eea
Furthermore, one can check that $\J_{\a(2l+2)\ad(2l+1)}$ satisfies the following
conservation equations:
\bea{l}
\D^{\a_{2l+2}}\J_{\a(2l+2)\ad(2l+1)}=0
~~,~~\Dd^{\a_{2l+1}}\J_{\a(2l+2)\ad(2l+1)}=0~~.\n\label{cce}
\eea
Expression (\ref{Jspc}) matches the Minkowski superspace limit of the AdS 
integer superspin supercurrents constructed in \cite{K4}. The above 
supercurrent is bilinear to the chiral superfield which describes free
massless fields of spin $0$ and $1/2$. In \cite{Cr} it was shown that
for such composite objects their proper transformation under conformal
symmetry is equivalent to the conservation equation conditions and the
on-shell equations of motion of its building blocks. Using similar 
arguments
one can confirm that $\J_{\a(2l+2)\ad(2l+1)}$ is a superconformal primary
with weights $(1+\tfrac{2l+2}{2}~,~1+\tfrac{2l+1}{2})$ which are
appropriate in order to make the
cubic interaction (\ref{cici}) superconformally invariant. This may feel
unexpected because our starting point, transformations (\ref{EdphiC}),
break conformal symmetry because they do not preserve the superconformal
primary property of $\Phi$. Nevertheless, the fact that we can
find an improvement term that makes the supertrace to vanish and at the same time generate a supercurrent with the appropriate conservation equations is a manifestation of the conformal symmetry of this 
special case. At the level of the action, the same special improvement terms that canceled the supertrace, recombined the effect of various terms in the transformation
in such a way in order to generate cubic interactions that
respect conformal symmetry. Obviously, this is not a general feature
and highly depends on the properties of the starting action. In this case,
$S_{0}$ has conformal symmetry, and for that reason we were able to
reach the \emph{minimal} supercurrent multiplet\footnote{This terminology was introduced in \cite{Ma} and we have used in \cite{BGK1,KKvU} in order to emphasize that these are special cases.} which resulted to a higher spin supercurrent which satisfies the correct conservation equations in order to be superconformally primary and thus restoring the conformal symmetry.

\section{Poincar\'e integer superspin supercurrents for free massless chiral}
\label{sec-Poincare}
Now, let's consider the effects of (\ref{EdphiP}) on the single, free, massless
chiral action (\ref{S0}). We get:
\bea{l}
\d_{g}S_{0}=g\sum_{k=0}^{\infty}\int\left[\vphantom{\frac12}
\D^2\ell^{\a(k+1)\ad(k)}
~\left\{(-i)^k\Phi~\pa^{(k)}\D\Phi\right\}\n\label{dSP1}
\right]+c.c.~~
\eea
As mentioned previously, the term inside the curly bracket has the property
$\Dd^{\ad_k}\left\{(-i)^k\Phi~\pa^{(k)}\D\Phi\right\}$=0, hence we can rewrite
(\ref{dSP1})
\bea{l}
\d_{g}S_{0}=g\sum_{k=1}^{\infty}\int\left[\vphantom{\frac12}
\left\{~\D^2\ell^{\a(k+1)\ad(k)}
~+~\tfrac{1}{k!}~\Dd^{(\ad_k}\lambda^{\a(k+1)\ad(k-1))}~\right\}
~\J_{\a(k+1)\ad(k)}~\right]\n\label{dSP}
+c.c.~~
\eea
where
\bea{l}
\J_{\a(k+1)\ad(k)}=(-i)^k\Phi~\pa^{(k)}\D\Phi~~.\n\label{JP}
\eea
Following Noether's method we find that this supercurrent generates the
following cubic interactions between free massless chiral supermultiplet and
the Poincar\'e integer superspin supermultiplet
\bea{l}
S_{I}=-g\sum_{s=2}^{\infty}\int~\Psi^{\a(s)\ad(s-1)}~\J_{\a(s)\ad(s-1)}
~+~c.c.\n\label{cipi}
\eea
In contrast with the previous conformal case, the supercurrent (and the cubic
interaction) is defined for every positive integer $s$ and not just for the even 
values. Furthermore, one can prove that $\J_{\a(s)\ad(s-1)}$ satisfy the 
following conservation equations:
\bea{l}
\D^2\J_{\a(s)\ad(s-1)}=0~~,~~\Dd^{\ad_{s-1}}\J_{\a(s)\ad(s-1)}=0\n\label{Pce}
\eea
and crucially $\D^{\a_{s}}\J_{\a(s)\ad(s-1)}\neq0$,
thus it can not be a primary superfield and it is not related with 
conformal supercurrent.

\section{Integer superspin supercurrent multiplet beyond free, massless, chiral}
In \cite{BGK2} we investigated the construction of half-integer superspin
supercurrent multiplet for a general class of non-linear sigma models
of a single chiral superfields, parametrized by an arbitrary K\"ahler potential 
$\K(\Phi,\bar{\Phi})$ and a chiral superpotential $\W(\Phi)$. The result was 
that besides the free, massless case, the arbitrary half-integer superspin 
supercurrent multiplets exist only for $\K(\Phi,\bar{\Phi})=\bar{\Phi}\Phi$ with 
$\W(\Phi)=f\Phi$ or $\W(\Phi)=m\Phi^2$, which is consistent
with the expectations coming from \cite{CM,HLS,MZ}.
Therefore, it will be interesting to
investigate the existence of arbitrary integer superspin supercurrents for the 
cases of a free chiral with a linear superpotential or a free massive chiral 
superfield. 

In both cases there is a dimension-full parameter, therefore only the
super-Poincar\'e higher spin supermultiplet is relevant. The general cubic 
interaction of the
chiral with the Poincar\'e integer superspin supermultiplet $Y=s$ (\ref{isst}) 
has the form
\bea{l}
S_{I}=\int d^8z\left[~\Psi^{\a(s)\ad(s-1)}~\J_{\a(s)\ad(s-1)}
~+~\tfrac{1}{2}~V^{\a(s-1)\ad(s-1)}~\T_{\a(s-1)\ad(s-1)}~\right]~+~c.c.\n
\eea
where $\J_{\a(s)\ad(s-1)}$ is the arbitrary integer superspin supercurrent and 
$\T_{\a(s-1)\ad(s-1)}$ is the arbitrary integer superspin supertrace
\footnote{The term supertrace originates from the cubic interaction of 
matter supermultiplets with the compensator of Poincar\'e supergravity (see
\cite{BK}). We have been using the same terminology for the higher spin version 
of this type of interactions, meaning the interactions with the compensator
of the Poincar\'e half-integer superspin supermultiplet. For the case of integer
superspin supermultiplets, we will continue to use it in the same spirit.
The supertrace generates the cubic interactions with the compensator of
the Poincar\'e integer superspin supermultiplet.}.
Due to the gauge symmetries (\ref{isst}) the supercurrent and the supertrace 
have to respect the following:
\bea{l}\n\label{fPce}
\D^2\J_{\a(s)\ad(s-1)}=\tfrac{1}{s!}~\D_{(\a_s}\T_{\a(s-1))\ad(s-1)}~~,\sn\\
\Dd^{\ad_{s-1}}\J_{\a(s)\ad(s-1)}=0\sn
\eea
where $\T_{\a(s-1)\ad(s-1)}$ is real.
In previous section, we demonstrated that for the free, massless chiral case the 
supertrace vanishes. However if we go beyond that, it is reasonable to expect
corrections proportional to the dimension-full parameter that controls the
added terms.
\subsection{Free chiral with linear superpotential}
Let's consider the addition of a linear superpotential term in (\ref{S0})
controlled by a complex parameter $f$:
\bea{l}
S_0=\int d^8z~ \bar{\Phi}\Phi ~+~f~\int d^6z \Phi 
~+~f^*~\int d^6\bar{z} \bar{\Phi}~.\n
\eea
It is straightforward to show that in this case the supercurrent and supertrace
are:
\bea{l}\n\label{JTlw}
\J_{\a(s)\ad(s-1)}=(-i)^{(s-1)}\Phi~\pa^{(s-1)}\D\Phi~~,\sn\\
\T_{\a(s-1)\ad(s-1)}=(-i)^{(s-1)}~f^*~\pa^{(s-1)}\Phi 
~+~ (i)^{(s-1)}~f~\pa^{(s-1)}\bar{\Phi}~.\sn 
\eea
The supercurrent does not acquire any modifications and remains the same
as in the massless case (\ref{JP}).
The supercurrent and supertrace can be defined for any ($s\geq2$) value of the 
integer $s$.
\subsection{Free massive chiral}
Let's consider the addition of a mass term in (\ref{S0})
\bea{l}
S_0=\int d^8z~ \bar{\Phi}\Phi ~+~m~\int d^6z \Phi^2
~+~m~\int d^6\bar{z} \bar{\Phi}^2\n
\eea
with a real mass parameter $m$.
In this case one can show that the integer superspin supercurrent and supertrace
exist only for even values of $s$ ($s=2,4,6,\dots$) and they are:
\bea{l}\n\label{JTqw}
\J_{\a(2l+2)\ad(2l+1)}=(-1)^{(l+1)}~i~\Phi~\pa^{(2l+1)}\D\Phi~~,\sn\\
\T_{\a(2l+1)\ad(2l+1)}=(-1)^{(l+1)}~i~m~\bar{\Phi}~\pa^{(2l+1)}\Phi 
~-~ (-1)^{(l+1)}~i~m~\Phi~\pa^{(2l+1)}\bar{\Phi}~~.\sn 
\eea

\section{Summary and discussion}
In recent work \cite{BGK1,KKvU} it has been shown that
\emph{linear}, higher spin transformations of matter superfields (such as the
chiral or the complex linear) exist and are responsible for generating 
(by following Noether's procedure) consistent cubic interactions between matter 
multiplets and \emph{half integer superspin} ($Y=s+1/2$) supermultiplets in 
terms of half-integer superspin supercurrents. Following this method we
were able to reproduce known supercurrents for a massless chiral and 
find new half-integer superspin supercurrents for a massive chiral. An 
interesting feature of these new supercurrents was the presence of a selection 
rule, meaning they can be defined only for odd values of the parameter $s$ 
($s=2l+1$).
These results were later reproduced in \cite{K2} following a different approach.

In this work we prove the existence of 
\emph{antilinear}, higher spin transformations of the chiral superfield which 
via Noether's method generate consistent cubic interactions between the chiral 
and the \emph{integer superspin} ($Y=s$) supermultiplets.
This is a very interesting feature because antilinear transformations do not 
appear frequently in physics. For linear transformations we have the
intuition of linearized superdiffeomorphisms that give rise to the cubic 
interactions of the theory to supergravity. However, we do not know an analogue 
example for antilinear transformations.
Using them, we were able to generate
new integer superspin supercurrents for the massless chiral and extend them to
the case of massive chiral and to the case of 
a linear superpotential. For some of these theories various selection rules 
emerge as well.

In detail, the work and results found in this paper are the following. We 
considered the most general, non-trivial, antilinear transformation of a 
chiral superfield which is consistent with the chiral constraint. In this case 
non-trivial means that we can not absorb the transformation by superfield 
redefinitions. For example, terms of the transformation that depend on the 
equations of motion (vanish on-shell) once they are introduced in the variation 
of the action can be ignored because they can be removed by trivial superfield 
redefinitions. This is equivalent with the usual argument of imposing the 
on-shell condition when we calculate the supercurrent. It is important to
realize
that such terms are necessary in order for the transformations to be chiral
but once they are used in order to calculate the variation of the action, they 
correspond 
to trivial redefinitions and do not contribute to the generation of non-trivial 
interactions. Thus they can be ignored from the very beginning. In this way we 
find two classes 
[(\ref{EdphiC}) and (\ref{EdphiP})] of effective transformations of this type.
Considering the effects of such transformations on the free massless chiral 
action we find:
\begin{enumerate}
\item[\emph{i}.~]Transformation (\ref{EdphiC}) leads to the construction of cubic 
interactions (\ref{cici}) with the conformal integer superspin supermultiplet
$Y=s$ but only for even values of $s$ ($s=2,4,\dots=2l+2$). The interactions are 
generated by the integer superspin supercurrent $\J_{\a(2l+2)\ad(2l+1)}$ given 
by (\ref{Jspc}), which satisfies conservation equations (\ref{cce}). This result
is consistent with the flat spacetime limit of the results in
\cite{K4} where the AdS
conformal integer superspin supercurrent was constructed by solving the 
conservation equation.
\item[\emph{ii}.~]Transformation (\ref{EdphiP}) leads to the construction of
cubic interactions (\ref{cipi}) with the Poincar\'e integer superspin
supermultiplet $Y=s$ for all values of $s$. The interactions are generated by
the supercurrent $\J_{\a(s)\ad(s-1)}$ given by (\ref{JP}) which satisfies the 
Poincar\'e conservation equations (\ref{Pce}) but not the conformal
conservation equations. This is a new supercurrent.
\end{enumerate}
Next, we considered matter theories beyond the simple free massless theory.
However based on \cite{CM,HLS,MZ,BGK2} the most general theory we can consider 
for the construction of higher spin supercurrents is the free, massless theory 
with the addition of a superpotential with linear and quadratic terms.
Due to the presence of dimension-full parameters that control these 
additional terms, we consider only the extention of the massless Poincar\'e
supercurrent. The approach here 
is to add all possible corrections to the previous Poincar\'e result and demand 
the conservation equations (\ref{fPce}) on-shell. The results we found are
the following new supercurrents:
\begin{enumerate}
\item[\emph{iii}.~]For a free chiral superfield with a linear superpotential,
we can construct cubic interactions with the Poincare integer superspin 
supermultiplet $Y=s$ for all values of $s$. The interactions are generated by
a supercurrent $\J_{\a(s)\ad(s-1)}$ and a supertrace $\T_{\a(s-1)\ad(s-1)}$
given by (\ref{JTlw}) with conservation equations (\ref{fPce}).
\item[\emph{iv}.~]For a free, massive chiral we find cubic interactions with the 
Poincar\'e integer superspin supermultiplet $Y=s$, but only for even values of
$s$ ($s=2,4,\dots=2l+2$). The supercurrent and supertrace that generate the 
interaction are given by (\ref{JTqw}).
\end{enumerate}
In this work, we considered chiral superfields to represent the matter
supermultiplets. However, similar constructions can be done for complex linear
superfields as is demonstrated in \cite{KKvU}. The fastest and easiest method to
extract the corresponding supercurrent multiplets for a complex linear 
superfield is to use the chiral - complex linear duality. Starting from
the supercurrents for the chiral and performing the duality one can get
the supercurrents for the complex linear as well as the relative coupling 
constant which relates the \emph{charge} of these two matter multiplets
for the interaction with higher superspins (see \cite{KKvU}). 
Additionally, it will be useful to comment that for any of the above higher spin
supercurrent multiplets, one can project the corresponding superspace  
conservation equations to components in order to find the usual spacetime
conservation equations and the corresponding higher spin current multiplets.
This has been illustrated in detail in \cite{BGK1,KKvU}.

Recently, the results of \cite{BGK1,KKvU} have been criticized in \cite{K4}
as incomplete.
For this reason we feel the need to clarify the results obtained
in \cite{BGK1,KKvU} in connection with the results obtained here.
In \cite{BGK1,KKvU} we considered only consistent, linear 
transformations of the matter supermultiplets and proved (via Noether's method) 
that they generate cubic interactions only with half-integer superspin 
supermultiplets. On the other hand, in 
\cite{K4}, using a different 
method (solving the superspace conservation equations) various
AdS cubic interactions have been constructed for both integer and half-integer 
superspin supermultiplets. The authors concluded that 
the results regarding integer superspin interactions have been overlooked in 
\cite{BGK1,KKvU}.
We communicated the context
of our work to the authors and we informed them 
that the interactions with the integer superspin
supermultiplets will originate from a consideration of antilinear 
transformations, as demonstrated in this work. An updated version
of \cite{K4}, included this argument and
an indicative expression of such antilinear transformations was added.

{\bf Acknowledgments}\\[.1in] \indent
The authors acknowledge S.\ M.\ Kuzenko
for the constructive criticism. The research of I.\ L.\ B.\ was supported in 
parts by Russian Ministry of Education and Science, project No. 3.1386.2017.
He is also grateful to RFBR grant, project No. 18-02-00153 for
partial support. The research of S.\ J.\ G.\ and K.\ K.\ is supported by the 
endowment of the Ford Foundation Professorship of Physics at 
Brown University. Also this work was partially supported by the U.S. National 
Science Foundation grant PHY-1315155. 

\end{document}